\documentclass[aps,twocolumn,amsmath,amssymb,floatfix,prl]{revtex4}
\usepackage{graphicx}
\usepackage{dcolumn}
\usepackage{bm}
\usepackage{amsfonts}
\usepackage{graphicx}
\usepackage{dcolumn}
\usepackage{bm}
\usepackage{amsmath}
\usepackage{amssymb}

\begin{document}

\title{Dimensionally mixed coupled collective modes} 
\author{E. H. Hwang$^{1,3}$, Ben Yu-Kuang Hu$^{2,3}$, and S. Das Sarma$^3$}
\affiliation{
$^1$SKKU Advanced Institute of Nanotechnology and Department of
Physics, Sungkyunkwan  University, Suwon, 16419, Korea \\
$^2$ Department of Physics, University of Akron, Akron, OH 44325 \\
$^3$Condensed Matter Theory Center and Joint Quantum Institute, 
Department of Physics, University of Maryland, College Park,
Maryland  20742-4111 
}



\begin{abstract}
We develop the diagrammatic formulation of the many-body theory for the coupled  collective modes in interacting
electron systems of different dimensions. The formalism is then applied in detail to a two-dimensional system coupled to a three-dimensional electron gas. 
We find two dimensionally-mixed plasmon modes which in the long wavelength limit are respectively 3D-like and acoustic in nature, but are strongly coupled at larger wave vectors.
The same formalism can be applied to any dimensional combinations, and we also present the results for 1D-2D and 1D-3D coupled systems. 
\end{abstract}

\maketitle



Collective modes (e.g., plasmons, magnons, phonons, zero sound, first sound, etc.) are manifest signatures of interacting many body systems, and are among the most-studied subjects in quantum many body theories.  In particular, understanding the plasmon modes of a three dimensional (3D) electron liquid interacting through the long range Coulomb interaction using several different theoretical approaches such as collective coordinates \cite{one}, dielectric function approach \cite{two}, diagrammatic perturbation theory \cite{three}, time dependent Hartree theory \cite{four}, and self-consistent equations of motion \cite{five} was the starting point of the modern many body theory for normal metals.  Similarly, studying the collective modes of normal He-3 \cite{six,seven} was among the key aspects of the modern development of the Fermi liquid theory. It was realized that understanding coupled collective modes of different systems leads to fundamental insights into the nature of finite many-body interacting systems, and as such, surface and interface plasmons (i.e., collective modes of metal-insulator systems) \cite{eight} and coupled electron-hole plasmons in interacting solid state plasmas \cite{nine} were also studied early in the history of many body theory.  In fact, an early result of studying coupled plasmon modes was the realization that the acoustic phonons in simple metals can be thought of equivalently as quantized sound waves or ionic plasmons screened by the conduction electrons in a coupled ion-electron quantum plasma.\cite{ten}  Later, both uncoupled and coupled plasmon modes of interacting 3D and two-dimensional  (2D) electron liquids were studied extensively both theoretically and experimentally in the context of doped semiconductor quantum wells and superlattices as well as 3D bulk systems.\cite{eleven,twelve,thirteen,thirteen1, thirteen2, fourteen, fifteen, sixteen, jain}  Collective modes of 1D electron systems have also been studied extensively \cite{seventeen, eighteen, nineteen, twenty} in the context of semiconductor quantum wire structures.

In spite of this great deal of activity encompassing thousands of publications on collective modes and plasmons in 3D, 2D, and one dimensional (1D) metallic systems as well as coupled collective modes in 3D-3D, 2D-2D, and 1D-1D interacting systems (going back to the early 1950s), there has not been much theoretical investigation of the nature of collective plasmon modes in systems of mixed dimensionality, i.e., 3D-2D or 3D-1D or 2D-1D systems.  This is surprising since it is possible to study such systems experimentally. For example, one can easily experimentally study the plasmon modes of a coupled system of a 2D electron layer close to a 3D electron slab or for that matter of a coupled system of a 2D layer and a 1D wire.  The goal of the current work is to develop a theoretical formalism to study this important and interesting topic:  What are the plasmon modes in coupled metallic  systems of mixed dimensionality? The answer to this question is not obvious at all.  First, the plasmon modes have completely different energy dispersions in different dimensions with the long wavelength plasma frequency going as $\sim q^{(3-D)/2}$ where D is the system dimensionality and $q$ is the relevant wave number.  Second, in systems of mixed dimensionality defining the wave number becomes ambiguous since different dimensions have wave vectors of different dimensions.

In this work we develop a general diagrammatic theory for obtaining the collective modes in coupled electronic systems of mixed dimensionality interacting through the long-range Coulomb interaction 
(i.e., the $1/r$ potential where $r$ is the physical distance between the two interacting electrons).  Our theory is applicable to systems of arbitrary dimensionality mixing.  We provide detailed results by applying our theory to the physically interesting coupled 3D-2D system.  We also give results for the coupled 2D-1D (and 3D-1D) system.

Our theory is based on obtaining the dynamically screened Coulomb interaction in the dimensionally coupled system by generalizing the appropriate 2-particle Dyson's equation for the interacting reducible polarizability.\cite{fetter}  We then proceed to solve this equation in the random phase approximation by approximating the irreducible polarizability by the corresponding noninteracting polarizability, which gives an exact result in the high-density limit of weak effective interaction, and this is also the exact classical result since the long-range Coulomb interaction is treated properly.
The poles of the reducible polarizability or the zeros of the corresponding dielectric function immediately give us the collective plasmon modes of the mixed system, which we proceed to obtain explicitly in the plasmon pole approximation which is exact to the leading order in wave vector.\cite{plasmonpole}

\begin{figure}[b]
\vspace{10pt}%
\includegraphics[width=1.\linewidth]{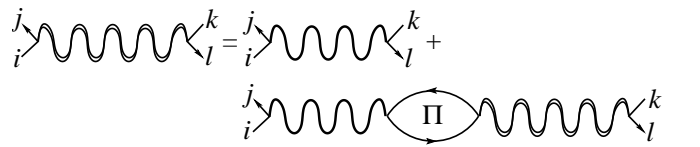}
\caption{Diagram for the screened Coulomb interaction. Here  $\Pi$ indicates the irreducible polarizability and single wiggled lines are the bare Coulomb interaction.}
\label{fig:fig1}
\end{figure}

Dyson's equation (Fig.~\ref{fig:fig1}) for the screened interaction ${\tilde W}$ in terms of the irreducible polarizability
$\Pi$ and the bare Coulomb interaction $V$ is, in general,
\begin{equation}
\tilde{W} = V + V \Pi \tilde{W},
\end{equation}
where the variables in this equation represent scattering matrices in the basis of two-particle
states since each interaction line corresponds to a scattering event between two electrons via the Coulomb interaction. 
For example, the matrix elements for $\tilde{W}$ are $\tilde{W}_{ij,kl} = \langle ik| \tilde{W} |jl \rangle$, representing scattering from states $|i\rangle \rightarrow |j\rangle$ for the first particle and
$| k \rangle \rightarrow | l \rangle$ for the second particle due to the screened interaction $\tilde{W}$. This equation is exact if $\Pi$ is the exact polarizability including the full interaction vertex.
In the event there are $N$ subsytems, which are coupled via interactions but without inter-system
tunneling, and we ignore terms in the $\Pi$ which connect different subsystems, then 
the Dyson equation becomes
\begin{equation}
\tilde{W}_{\alpha\beta} = V_{\alpha\beta} + \sum_{\gamma=1}^{N}V_{\alpha\gamma} \Pi_{\gamma}^0 \tilde{W}_{\alpha\beta},
\end{equation}
where $\alpha,\;\beta=1,...,N$.
Here $\tilde{W}_{\alpha\beta}$ represent scattering matrices where the eigenstates are localized 
within subsystems $\alpha$ and $\beta$ (e.g., $\tilde{W}_{i_{\alpha}j_{\alpha},k_{\beta}l_{\beta}}$ 
where $i_{\alpha}$ and $j_{\alpha}$ are states in system $\alpha$, and $k_{\beta}$ and $l_{\beta}$ are states in system $\beta$)  and $\Pi_{\gamma}^0$ is the irreducible polarizability of subsystem $\gamma$. We note that our no inter-subsystem tunneling approximation leads to a particle conservation in each individual subsystem, which is a reasonable assumption.

We now apply the theory to a system which consists of two subsystems, a 2D quantum well of zero
width in the x-y plane of area A (system 1) embedded in a bulk system of 3D conduction electrons of
length $L_z$ in the $z$-direction (system 2). We use label ``1" for a 2D system and label ``2" for a 3D system. The Hamiltonian of the total system is
\begin{equation}
H=H_1+H_2 + H_{12},
\end{equation}
where $H_{\alpha}$ is the Hamiltonian for the isolated system $\alpha=1,2$, and $H_{12}$ represents the interaction between two systems
\begin{eqnarray}
H_{12} = \sum_{{\bf k}_1,{\bf k}_2,{\bf q}}\sum_{k_z,q_z}V_{12}({\bf q},q_z)c_2^{\dagger}({\bf k}_1-{\bf q},k_z-q_z)c_1^{\dagger} \nonumber \\
\times ({\bf k}_1+{\bf q})c_1({\bf k}_1)c_2({\bf k}_2,k_z),
\end{eqnarray}
where $c_{\alpha}^{\dagger}$ ($c_{\alpha}$) is creation
(destruction) operators in system $\alpha$ and $V_{12}$ is the Coulomb potential between the two systems.

The equations needed for the screened Coulomb interaction between electrons in the subsystem 1, $\tilde{W}_{11}$, are (suppressing the dependence on wave number $q$ and frequency $\omega$, which all of these depend on)
\begin{equation}
\tilde{W}_{11}=V_{11} + V_{11}\Pi_1^0 \tilde{W}_{11} + \sum_{q_z}V_{12}\Pi_2^0\tilde{W}_{21},
\end{equation}
\begin{equation}
\tilde{W}_{21}=V_{21}+V_{21}\Pi_1^0\tilde{W}_{11}+V_{22}\Pi_2^0\tilde{W}_{21},
\end{equation}
where $\Pi_{\alpha}^0$ is the irreducible polarizability in subsystem $\alpha = 1, 2$. These equations give, reintroducing the $q = |{\bf q}|$ and $\omega$ dependence (where {\bf q} is the two-dimensional wave vector)
\begin{equation}
\tilde{W}_{11}(q,\omega)=\frac{V_{11}^{\rm eff}(q,\omega)}{\epsilon_1^{\rm eff}(q,\omega)},
\end{equation}
where $V_{11}^{\rm eff}$ and $\epsilon_1^{\rm eff}$ are given by, respectively,
\begin{equation}
V_{11}^{\rm eff}(q,\omega) = V_{11}(q) + \int \frac{d q_z}{2\pi} |V_{12}(Q)|^2 \frac{\Pi_2^0(Q,\omega)}{\epsilon_2(Q,\omega)},
\end{equation}
\begin{equation}
\epsilon_1^{\rm eff}(q,\omega) = 1- V_{11}^{\rm eff}(q,\omega) \Pi_1^0(q,\omega),
\end{equation}
where $Q=\sqrt{q^2 + q_z^2}$.
The dielectric functions for subsystem 1 and 2 are given by 
\begin{equation}
\epsilon_1(q,\omega) = 1-V_{11}(q)\Pi_1^0(q,\omega),
\end{equation}
\begin{equation}
\epsilon_2(Q,\omega) = 1-V_{22}(Q)\Pi_2^0(Q,\omega),
\end{equation}
and the Coulomb interactions are given by
\begin{equation}
V_{11}(q) = \frac{2\pi e^2}{q},
\label{2dcoulomb}
\end{equation}
\begin{equation}
V_{22}(Q) = \frac{4\pi e^2}{Q^2},
\label{3dcoulomb}
\end{equation}
and the interaction between the two subsystems 
\begin{equation}
V_{12}({\bf q},q_z) = \frac{4\pi e^2}{Q^2}e^{-qd},
\label{23dcoulomb}
\end{equation}
where $d$ is the separation distance between the two systems.
We note that Eqs.~(\ref{2dcoulomb}), (\ref{3dcoulomb}), and (\ref{23dcoulomb}) give respectively the Fourier transforms of the usual $1/r$ Coulomb interaction in pure 3D, pure 2D, and 3D-2D mixed dimensions

We also find the screened Coulomb interaction of the subsystem 2, $\tilde{W}_{2}$, as
\begin{equation}
\tilde{W}_{22}(Q,\omega)=\frac{V_{22}^{\rm eff}(Q,\omega)}{\epsilon_2^{\rm eff}(Q,\omega)},
\end{equation}
where $V_{22}^{\rm eff}$ and $\epsilon_2^{\rm eff}$ are given by, respectively,
\begin{equation}
V_{22}^{\rm eff}(Q,\omega) = V_{22}(Q) + V_{22}(Q) \frac{V_{11}(q)e^{-2qd}\Pi_1^0(q,\omega)}{\epsilon_1(q,\omega)},
\end{equation}
\begin{equation}
\epsilon_2^{\rm eff}(Q,\omega) = 1- V_{22}^{\rm eff}(Q,\omega) \Pi_2^0(Q,\omega).
\label{epsilon3}
\end{equation}
The collective modes of the systems now can be calculated by finding zeroes of $\epsilon_{\alpha}^{\rm eff}(q,\omega)$, i.e.,
\begin{equation}
\epsilon_{\alpha}^{\rm eff}(q,\omega) = 0.
\label{epsilon_alpha}
\end{equation}

The irreducible polarizability function $\Pi^0$ of an interacting electron system is unknown since self-energy and vertex corrections cannot be calculated exactly. 
However, in the long wavelength limit ($q \rightarrow 0$), the dielectric function, and, consequently, the plasmon frequency is determined entirely by the noninteracting irreducible polarizability, i.e., the electron-hole bubble diagram. 
For the Coulomb interacting system, this also gives the exact result asymptotically in the high-density limit where all interaction corrections to the irreducible polarizability vanish in a well-defined series expansion manner.\cite{fetter}

The noninteracting irreducible polarizability in D-dimension is given by the expression with D-dimensional wave vectors $k$ and $q_D$ \cite{hwang2009}
\begin{equation}
\Pi^0(q_D,\omega)=g \int \frac{d^Dk}{(2\pi)^D} \frac{n_F(\xi_k)-n_F(\xi_{k+q_D})} {\omega+\xi_k-\xi_{k+q_D}},
\end{equation}
where $g$ is the degeneracy factor (e.g., spin, valley), $\xi_k$ is the single-particle energy dispersion measured from the Fermi energy, i.e., 
$\xi_k=k^2/2m - E_F$, and $n_F$ is the Fermi distribution function. 
The leading order behavior of the noninteracting irreducible polarizability in the longwavelength limit, $q_D/\omega \ll 1$, can be obtained as
\begin{equation}
\Pi^0(q_D,\omega) = \frac{n_D}{m}\frac{q_D^2}{\omega^2} + O(q^4_D/\omega^4),
\end{equation}
where $n_D$ is the D-dimensional electron density.
With this result and Eq.~(\ref{epsilon_alpha}) we find two coupled plasmon modes in the long wavelength limit as
\begin{equation}
\omega_+ (Q) = \sqrt{ (\omega^{3D}_Q)^2 + (\omega^{2D}_q)^2 },
\label{wplus}
\end{equation}
\begin{equation}
\omega_- (Q)=  \frac{\omega_Q^{3D} \omega_q^{2D}} {\sqrt{(\omega^{3D}_Q)^2 + (\omega^{2D}_q)^2} }
\sqrt{ 1- e^{-2qd} },
\label{wminus}
\end{equation}
where $\omega_Q^{3D}$ and $\omega_q^{2D}$ are the long wavelength plasma frequency for the 3D and 2D electron gases (with Q and q being the 3D and 2D wavenumbers),
respectively, which are given by \cite{fourteen,fetter}
\begin{equation}
\omega_Q^{3D} = \sqrt{\frac{4\pi n_{3}e^2}{m}} + O(Q^2),
\label{w3d}
\end{equation}
\begin{equation}
\omega_q^{2D} = \sqrt{2\pi n_{2} e^2/m} \sqrt{q}.
\end{equation}

\begin{figure}[b]
\vspace{10pt}%
\includegraphics[width=0.9\linewidth]{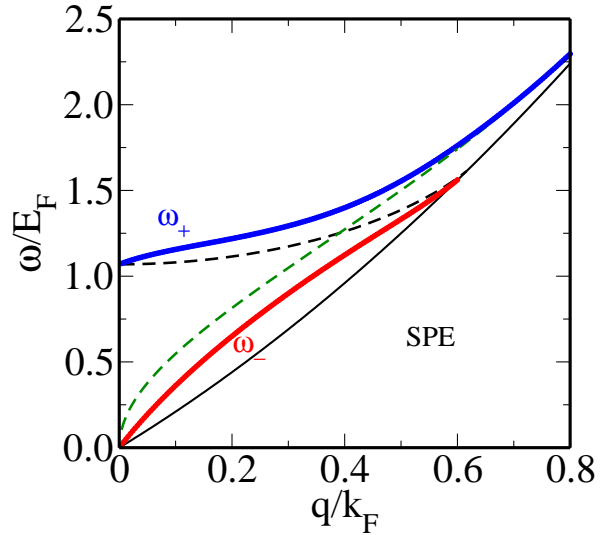}
\caption{
Calculated plasmon dispersions of the 2D-3D coupled systems as a function of the wave vector $Q=q$ with $q_z=0$. Here we use the parameters corresponding to $k_F^{2D} = k_F^{3D}$, which indicates $n_3 = (2\pi n_2)^{3/2}/(3\pi^2)$. $d=5$nm, the separation distance between two systems, is used in this calculation. Both modes increase linearly in the long wavelength limit.
The dashed lines represent the uncoupled 3D (black) and 2D (green) plasmon modes, and the solid black line show the boundary of the single particle excitation region.}
\label{fig:polarization_function}
\end{figure}

\begin{figure}[htb]
\vspace{10pt}%
\includegraphics[width=0.9\linewidth]{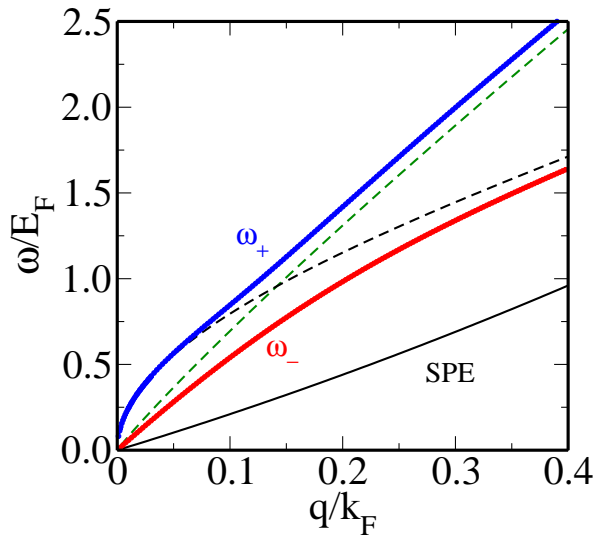}
\caption{
Calculated plasmon dispersions of the 1D-2D coupled systems as a function of the wave vector $q=q_x$ with $q_y=0$. Here we use the parameters corresponding to $k_F^{1D} = k_F^{2D}$, which indicates $n_2 = (\pi/8) n_1^2$. $d=4$nm, the separation distance between two systems, is used in this calculation. The dashed lines represent the uncoupled 2D (black) and 1D (green) plasmon modes, and the solid black line show the boundary of the single particle excitation region.}
\label{fig:polarization_function}
\end{figure}

From the above results we find several interesting features of the coupled plasmon modes:
(1) The leading order of the plasmon energy dispersions is independent of $q_z$. (2) $\omega_+$ mode has exactly the same energy as the 3D plasmon at $Q=0$, and the leading order correction to $\omega_+$ is linear in wave vector. In general, the quantum 
wave vector dispersion
correction of the 3D plasma frequency is quadratic,\cite{fetter} but here the dispersion correction to the $\omega_+$ mode is linear in wave vector in contrast to the corresponding pure 3D situation.
(3) $\omega_-(q)$ mode increases linearly with wave vector, i.e., $\omega_-(q) \propto q$.  Thus in 2D-3D coupled system there is no mode behaving like a 2D plasmon where $\omega (q) \propto \sqrt{q}$. Instead, the higher-energy  mixed plasmon $\omega_+$ has the long-wavelength 3D plasma frequency, but with a linear wavevector dispersion correction whereas the lower energy $\omega_-$ mixed mode has a purely acoustic energy dispersion going linear in wave vector.

In Fig.~2 we provide the numerical results for our calculated 2D-3D plasmons based on Eq.~(\ref{epsilon_alpha}) without making any long wavelength approximation.  The long wavelength coupled collective mode dispersions precisely obey our analytical results for $\omega_+$ and $\omega_-$ given in Eqs.~(\ref{wplus}) and (\ref{wminus}), but there is strong mode repulsion effect at larger wavevectors which cannot be captured in the long wavelength analytical approximations.

We can apply the same methods for a 1D-2D coupled system as we have developed above for the 2D-3D system. But due to the form factor which arises from a finite width of the 1D system
(in a strict zero width 1D system, the Coulomb interaction diverges logarithmically in momentum space),
the Coulomb potential of a 1D system is logarithmically
dependent on the width of the 1D wire. \cite{seventeen}
Now we take the 2D layer to be in the $x$-$y$ plane at $z=0$ and the 1D wire with finite width $a$ to be along the $x$-axis at the $y=0$ and $z=d$ plane. Then we can consider the strict long wavelength limit along $y$-direction (i.e. the system wavenumber being strictly along the $x$ direction by construction).  Now $q$ means $q_x$ along the $x$ direction which is conserved.  With the above geometry for the coupled 1D-2D
system we get the following results at long wavelength, $q\rightarrow 0$,
\begin{equation}
\omega_+ (q) = \sqrt{ (\omega^{2D}_q)^2 + (\omega^{1D}_{q})^2 },
\label{wplus1d}
\end{equation}
\begin{equation}
\omega_- (q)=  \frac{\omega_q^{2D} \omega_{q}^{1D}} {\sqrt{(\omega^{2D}_q)^2 + (\omega^{1D}_{q})^2V_q }} \sqrt{ V_q- I_q },
\label{wminus1d}
\end{equation}  
where 
\begin{equation}
\omega_q^{1D} = \sqrt{\frac{2n_1e^2}{m}} q,
\end{equation}
and $V_q$, $I_q$ arise from the finite width of the 1D wire, \cite{seventeen,nineteen}
and they depend on the specific quantum wire systems. If we take a quantum wire with an infinite square well potential confinement then we get 
\begin{equation}
V_q \sim C_1 + |\ln (qa)|,
\end{equation}
and
\begin{eqnarray}
I_q & \sim &  \left [ C_2 + |\ln (qa)| \right] \exp (-2qd) \;\;\;\; {\rm for \;\; d\ll a}, \nonumber \\
& \sim & \ln(a/d) + |\ln(qa)| \;\;\;\;\;\;\; {\rm for \;\; d \gg a},
\end{eqnarray}
where $C_1 = 2.08862$ and $C_2 = 0.922784$ are constants.
Since $\ln(qa)$ term exactly cancels in $V_q-I_q$, the acoustic mode increases linearly  (i.e., $\omega_-(q) \propto q$) without logarithmic correction in the long wavelength limit,\cite{seventeen} but the slope is dependent on the ratio of $d/a$. 
Note that the situation for the 1D-2D coupled mixed plasmons is qualitatively similar to the 2D-3D plasmon case, i.e., the $\omega_+$ mode [Eq.~(\ref{wplus1d})] follows the 2D plasmon dispersion of $q^{1/2}$ in the long wavelength with the first order correction being linear in $q$ whereas the $\omega_-$ mode [Eq.~(\ref{wminus1d})] is acoustic in nature, going simply as linear in q at long wavelength.

In Fig. 3, we provide our numerically calculated 1D-2D coupled plasmon dispersion, which agrees with our analytical results at long wavelength, but manifests mode repulsion effects at intermediate wavelength.

Our theory can also be applied to the mixed 1D-3D plasmon geometry, where the coupled modes are easily found in the leading order to obey the dispersion relations: $\omega_+ = \omega_q^{3D} + O(q^2)$ and $\omega_- \sim O(q)$, where $\omega_q^{3D}$, given by the first term in Eq.~(\ref{w3d}) is the standard long wavelength 3D plasma frequency and $q$ is the wave number along the 1D system.  The $O(q^2)$ nonlocal dispersion term in the $\omega_+$ mode contains parameters corresponding to both the 3D and the 1D system as well as the spatial separation '$d$' between them.
The slope of the acoustic plasmon mode $\omega_-$ also depends on the distance `$d$' between the 1D and the 3D system.

In summary, we have developed a theory for coupled plasmons in mixed dimensionality and applied it to obtain the collective modes in mixed 2D-3D, 1D-2D, and 1D-3D metallic systems.  The two coupled modes of the system manifest interesting long wavelength dispersion with the higher mode always showing the collective plasma frequency of the relative higher dimensional system (i.e., 3D plasmon for the 2D-3D and 1D-3D cases, and 2D plasmon for the 1D-2D case) and the lower plasmon mode manifesting acoustic dispersion linear in wave vector.  The leading order dispersion corrections to the higher energy mode is linear in the wave vector in both 2D-3D and 1D-2D situations, but is quadratic in the 1D-3D case.
The basic physics of dimensionally mixed coupled collective modes is that the higher dimensional system always screens out the lower dimensional collective mode into an acoustic mode whereas the leading term in the higher dimensional plasmon remains intact with only the nonlocal dispersion corrections being affected by the mode coupling.  It should be straightforward to experimentally verify our predictions through inelastic electron energy loss, optical far infrared, and inelastic light scattering spectroscopies in doped 1D, 2D, 3D semiconductor systems.


This work is supported by the Laboratory for Physical Sciences.
EHH also acknowledges support from  Basic Science Research Program (2017R1A2A2A05001403) of the National Research Foundation of Korea.




\begin{thebibliography}{999}

\bibitem{one} D. Bohm and D. Pines, Phys. Rev. {\bf 92}, 609 (1953).

\bibitem{two} J. Lindhard, K. Dan. Vidensk. Selsk. Mat. Fys. Medd. {\bf 28}, No. 8 (1954).

\bibitem{three} M. Gell-Mann and K. A. Brueckner, Phys. Rev. {\bf 106}, 364 (1957).

\bibitem{four} H. Ehrenreich and M. H. Cohen, Phys. Rev. {\bf 115}, 786 (1959). 

\bibitem{five} K. S. Singwi, M. P. Tosi, R. H. Land, and A. Sj\"{o}lander, Phys. Rev. {\bf 176}, 589 (1968).

\bibitem{six} G. Baym and C. Pethick, {\it Landau Fermi Liquid Theory: Concepts and Applications} (Wiley-VCH Verlag, 1991).

\bibitem{seven} D. Pines and P. Nozieres, {\it The Theory of Quantum Liquids} (W. A. Benjamin, New York, 1966).

\bibitem{eight} R. H. Ritchie, Phys. Rev. {\bf 106}, 874 (1957); 
E. A. Stern and R. A. Ferrell, Phys. Rev. {\bf 120}, 130 (1960).

\bibitem{nine} D. Pines and J. R. Schrieffer, Phys. Rev. {\bf 124}, 1387(1961); {\it ibid} {\bf 125}, 804 (1962); D. Pines, Can. J. Phys. {\bf 34}, 1379 (1956).

\bibitem{ten} J. Bardeen and D. Pines, Phys. Rev. {\bf 99}, 1140 (1955).

\bibitem{eleven} S. Das Sarma and A. Madhukar, Phys. Rev. B {\bf 23}, 805 (1981).

\bibitem{twelve} S. Das Sarma and J.J. Quinn, Phys. Rev. B {\bf 25}, 7603 (1982).

\bibitem{thirteen} D. Olego, A. Pinczuk, A. C. Gossard, and W. Wiegmann,
Phys. Rev. B {\bf 25}, 7867 (1982); A. Pinczuk, J. Shah, and P. A. Wolff,
Phys. Rev. Lett. {\bf 47}, 1487 (1981);
A. Pinczuk, M. G. Lamont, and A. C. Gossard, Phys. Rev. Lett. {\bf 56}, 2092 (1986).

\bibitem{thirteen1} G. Fasol, N. Mestres, H. P. Hughes, A. Fischer, and K. Ploog,
Phys. Rev. Lett. {\bf 56}, 2517 (1986). 

\bibitem{thirteen2} A. Mooradian and G. B. Wright, Phys. Rev. Lett. {\bf 16}, 999 (1966);
A. Mooradian and A. L.  McWhorter, {\it ibid} {\bf 19}, 849 (1967).

\bibitem{fourteen} F. Stern, Phys. Rev. Lett. {\bf 18}, 546 (1967).

\bibitem{fifteen} S. J. Allen, Jr., D. C. Tsui, and R. A. Logan, Phys. Rev. Lett. {\bf 38}, 980 (1977).

\bibitem{sixteen} T. Ando, A. B. Fowler, and F. Stern, Rev. Mod. Phys. {\bf 54}, 437 (1982).

\bibitem{jain} J. K. Jain and P. B. Allen, Phys. Rev. Lett. {\bf 54}, 1985  (1985).

\bibitem{seventeen} S. Das Sarma and Wu-yan Lai, Phys. Rev. B {\bf 32}, 1401 (1985).

\bibitem{eighteen} Q. P. Li, S. Das Sarma, and R. Joynt, Phys. Rev. B {\bf 45}, 13713 (1992).


\bibitem{nineteen} S. Das Sarma and E. H. Hwang, Phys. Rev. B {\bf 54}, 1936 (1996);
 {\it ibid} {\bf 59}, 10730 (1999).


\bibitem{twenty} 
A. R. Goni, A. Pinczuk, J. S. Weiner, J. M. Calleja, B. S. Dennis, L. N. Pfeiffer, and K. W. West,
Phys. Rev. Lett. {\bf 67}, 3298 (1991);



\bibitem{fetter} A. L. Fetter and J. D. Walecka, {\it Quantum Theory of Many-particle Systems}
(McGraw-Hill, New York, 1971);
P. M. Platzmann and P. A. Wolff, {\it Waves and Interactions in Solid State Plasmas, Solid State Physics} (Academic Press, New York, 1973);
D. Pines, {\it Elementary excitations in solids} (Perseus Books, Reading, 1999);
A. A. Abrikosov, L. P. Gorkov, and I. E. Dzyaloshinskii, {\it Methods of Quantum Field Theory in Statistical Physics} (Dover, New York, 1975).

\bibitem{plasmonpole} B. I. Lundqvist, Phys. Konden. Mater. {\bf 6}, 193 (1967); B. Vinter, Phys. Rev. B {\bf 13}, 4447 (1976); S. Das Sarma, E. H. Hwang, and L. Zheng, Phys. Rev. B {\bf 54}, 8057 (1996).

\bibitem{hwang2009} S. Das Sarma and E. H. Hwang, Phys. Rev. Lett. {\bf 102}, 206412 (2009).

\end{thebibliography}
\end{document}